\documentstyle[epsfig]{aipproc2}
\begin{document}
%==========================%
%<<<<<      Title     >>>>>%
%==========================%
\title{Entanglement/Brick-wall entropies correspondence}
\author{Shinji Mukohyama}
\address{Department of Physics and Astronomy, 
University of Victoria, Victoria, BC, Canada V8W 3P6 \\
Yukawa Institute for Theoretical Physics, 
Kyoto University, Kyoto 606-8502, Japan}
\maketitle

%==========================%
%<<<<<    Abstract    >>>>>%
%==========================%
\begin{abstract}
There have been many attempts to understand the statistical origin of
black-hole entropy. Among them, entanglement entropy and the brick
wall model are strong candidates.
In this paper we show a relation between entanglement entropy and the
brick wall model: the brick wall model seeks the maximal value of the
entanglement entropy. 
In other words, the entanglement approach reduces to the brick wall
model when we seek the maximal entanglement entropy .
\end{abstract}

%==========================%
%<<<<<  Introduction  >>>>>%
%==========================%
\section{Introduction}

Black hole entropy is given by a mysterious formula called the
Bekenstein-Hawking formula~\cite{Bekenstein,Hawking}: 
%============< EQUATION >==============%
%
\begin{equation}
 S_{BH} = \frac{A}{4l_{pl}^2},
\end{equation}
%======================================%
where $A$ is area of the horizon. There have been many attempts to
understand the statistical origin of the black-hole entropy.

Entanglement entropy~\cite{BKLS,Srednicki} is one of the strongest
candidates of the origin of black hole entropy. It is originated from
a direct-sum structure of a Hilbert space of a quantum system: for an
element $|\psi\rangle$ of the Hilbert space ${\cal F}$ of the form 
%============< EQUATION >==============%
%
\begin{equation}
 {\cal F} = {\cal F}_I \bar{\otimes} {\cal F}_{II},
	\label{eqn:F=F1*F2}
\end{equation}
%======================================%
the entanglement entropy $S_{ent}$ is defined by 
%============< EQUATION >==============%
%
\begin{eqnarray}
 S_{ent} & = & -{\bf Tr}_I[\rho_I\ln\rho_I],\nonumber\\
 \rho_I  & = & {\bf Tr}_{II}|\psi\rangle\langle\psi|.
\end{eqnarray}
%======================================%
Here $\bar{\otimes}$ denotes a tensor product followed by a suitable
completion and ${\bf Tr}_{I,II}$ denotes a partial trace over 
${\cal F}_{I,II}$, respectively.

On the other hand, there is another strong candidate for the origin of 
black hole entropy: the brick wall model introduced by
'tHooft~\cite{tHooft}. In this model, thermal atmosphere in
equilibrium with a black hole is considered. In this situation, we
encounter with two kinds of divergences in physical quantities. The
first is due to infinite volume of the system and the second is due
to infinite blue shift near the horizon. We are not interested in the
first since it represents contribution from matter in the far
distance. Hence we introduce an outer boundary in order to make our
system finite. It is the second divergence that we would like to
associate with black hole entropy. Namely, it can be shown by
introducing a Planck scale cutoff that entropy of the thermal
atmosphere near the horizon is proportional to the area of the horizon
in Planck units.

In this paper, we show that the brick wall model seeks the maximal
value of the entanglement entropy.

%=====================================%
%<<<<<     Model description     >>>>>%
%=====================================%
\section{Model description}

\begin{figure}
\centerline{\epsfig{file=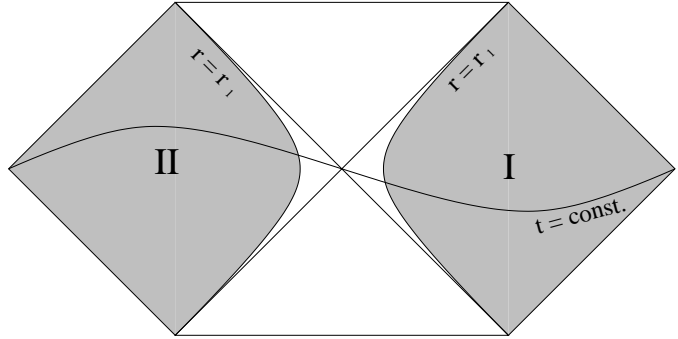,width=3.5in}}
\caption{The Kruskal-like extension of the static, spherically
symmetric black-hole spacetime. We consider only the regions
satisfying $r>r_1$ (the shaded regions $I$ and $II$ ).}
\vspace*{10pt}
\label{fig:Kruskal}
\end{figure}

For simplicity, we consider a minimally coupled, real scalar field
described by the action 
%============< EQUATION >==============%
%
\begin{equation}
 S = -\frac{1}{2}\int d^4x\sqrt{-g}\left[
        g^{\mu\nu}\partial_{\mu}\phi\partial_{\nu}\phi
        + m_{\phi}^2\phi^2\right], 
\end{equation}
%======================================%
in the spherically symmetric, static black-hole spacetime
%============< EQUATION >==============%
%
\begin{equation}
 ds^2 = -f(r)dt^2 + \frac{dr^2}{f(r)} + r^2d\Omega^2.
\end{equation}
%======================================%
We denote the area radius of the horizon by $r_0$ and the surface
gravity by $\kappa_0$ ($\ne 0$):
%============< EQUATION >==============%
%
\begin{eqnarray}
 f(r_0) & = & 0,\nonumber\\
 \kappa_0 & = & \frac{1}{2}f'(r_0).
\end{eqnarray}
%======================================%
We quantize the system of the scalar field with respect to the Killing 
time $t$ in a Kruskal-like extension of the black hole spacetime. 
The corresponding ground state is called the Boulware state and its 
energy density is known to diverge near the horizon. 
Although we shall only consider states with bounded energy density, it
is convenient to express these states as excited states above the
Boulware ground state for technical reasons. 
Hence, we would like to introduce an ultraviolet cutoff $\alpha$ with 
dimension of length to control the divergence. 
The cutoff parameter $\alpha$ is implemented so that we only consider
two regions satisfying $r>r_1$ (shaded regions $I$ and $II$ in 
{\it Figure}~\ref{fig:Kruskal}), where $r_1$ ($>r_0$) is determined by 
%============< EQUATION >==============%
%
\begin{equation}
 \alpha = \int_{r_0}^{r_1}\frac{dr}{\sqrt{f(r)}}. 
\end{equation}
%======================================%
[Evidently, the limit $\alpha\to 0$ corresponds to the limit 
$r_1\to r_0$. Thus, in this limit, the whole region in which 
$\partial /\partial t$ is timelike is considered.]
Strictly speaking, we also have to introduce outer boundaries, say at 
$r=L$ ($\gg r_0$), to control the infinite volume of the constant-$t$
surface. However, even if there are outer boundaries, the
following arguments still hold.

%============< FIGURE >==============%
%              Kruskal.ps
%======================================%

In this situation, there is a natural choice for division of the
system of the scalar field: let ${\cal H}_I$ be the space of mode
functions with supports in the region $I$ and ${\cal H}_{II}$ be the 
space of mode functions with supports in the region $II$. 
Thence, the space ${\cal F}$ of all states are of the form
(\ref{eqn:F=F1*F2}), where ${\cal F}_I$ and ${\cal F}_{II}$ are
defined as symmetric Fock spaces constructed from ${\cal H}_I$ and
${\cal H}_{II}$, respectively:
%============< EQUATION >==============%
%
\begin{eqnarray}
 {\cal F}_I & \equiv &
    \mbox{\boldmath C} \oplus {\cal H}_I \oplus 
    \left({\cal H}_I\bar{\otimes} {\cal H}_I\right)_{sym}
    \oplus \cdots,      
    \nonumber   \\
 {\cal F}_{II} & \equiv &  
    \mbox{\boldmath C} \oplus {\cal H}_{II} \oplus 
    \left({\cal H}_{II} \bar{\otimes} {\cal H}_{II} \right)_{sym} 
    \oplus \cdots.
\end{eqnarray}
%======================================%
Here $(\cdots)_{sym}$ denotes the symmetrization.

%============================================%
%<<<<<   Small backreaction condition   >>>>>%
%============================================%
\section{Small backreaction condition}

Let us investigate what kind of condition should be imposed for our
arguments to be self-consistent. 
A clear condition is that the backreaction of the scalar field to the
background geometry should be finite.
For the brick wall model this condition is satisfied. Namely, in
Ref.~\cite{Mukohyama&Israel}, it was shown that the total mass of the
thermal atmosphere of quantum fields is actually bounded. 
Thus, also for our system, we would like to impose the condition that 
the contribution $\Delta M$ of the subsystem ${\cal F}_I$ to the mass
of the background geometry should be bounded in the limit 
$\alpha\to 0$.

It is easily shown that $\Delta M$ is given by
%============< EQUATION >==============%
%
\begin{equation}
 \Delta M \equiv 
	-\int_{x\in I}T^t_t 4\pi r^2dr = H_I,
\end{equation}
%======================================%
where $H_I$ is the Hamiltonian of the subsystem ${\cal F}_I$ with
respect to the Killing time $t$.
Hence, the expectation value of $\Delta M$ with respect to a state
$|\psi\rangle$ of the scalar field is decomposed into the contribution 
of excitations and the contribution from the zero-point energy: 
%============< EQUATION >==============%
%
\begin{equation}
 \langle\psi |\Delta M|\psi\rangle = E_{ent} + \Delta M_{B},
\end{equation}
%======================================%
where $E_{ent}$ is entanglement energy defined by
%============< EQUATION >==============%
%
\begin{equation}
 E_{ent} \equiv \langle\psi |:H_I:|\psi\rangle, \label{eqn:Eent}
\end{equation}
%======================================%
and $\Delta M_{B}$ is the zero-point energy of the Boulware state. 
Here, the colons denote the usual normal ordering. 
[This definition of entanglement energy corresponds to
$E_{ent}^{(I')}$ in Ref.~\cite{MSK1998} and $\langle :H_2:\rangle$ in
Ref.~\cite{D-thesis}.]

Since the Boulware energy $\Delta M_{B}$ diverges as 
$\Delta M_{B}\sim -AT_H\alpha^{-2}$ in the limit 
$\alpha\to 0$~\cite{Mukohyama&Israel}, we should impose the condition 
%============< EQUATION >==============%
%
\begin{equation}
 E_{ent} \simeq |\Delta M_B|, \label{eqn:SBC}
\end{equation}
%======================================%
where $A=4\pi r_0^2$ is the area of the horizon, $T_H=\kappa_0/2\pi$
is the Hawking temperature. We would like to call this condition 
{\it the small backreaction condition (SBC)}. Note that the right hand 
side of SBC (\ref{eqn:SBC}) is independent of the state
$|\psi\rangle$.

%==========================================================%
%<<<<<   State having Maximal entanglement entropy    >>>>>%
%==========================================================%
\section{Maximal entanglement entropy}

Now, we shall show that the Hartle-Hawking state is a maximum of the 
entanglement entropy in the space of quantum states satisfying SBC. 
For this purpose, we prove a more general statement for a quantum
system with a state-space of the form (\ref{eqn:F=F1*F2}): 
{\it a state of the form 
%============< EQUATION >==============%
%
\begin{equation}
 |\psi\rangle = {\cal N} \sum_n 
	e^{-E_n/2T}|n\rangle_{I}\otimes|n\rangle_{II}
	\label{eqn:HH-state}
\end{equation}
%======================================%
is a maximum of the entanglement entropy in the space of states with
fixed expectation value of the operator $E_{I}$ defined by 
%============< EQUATION >==============%
%
\begin{equation}
 E_{I} = \left(\sum_n E_n|n\rangle_{I}\cdot{}_{I}\langle n|\right)
	\otimes 
	\left(\sum_m |m\rangle_{II}\cdot{}_{II}\langle m|\right),
\end{equation}
%======================================%
provided that the real constant $T$ is determined so that the
expectation value of $E_I$ is actually the fixed value. Here,
$\{|n\rangle_I\}$ and $\{|n\rangle_{II}\}$ ($n=1,2,\cdots$) are bases
of the subspaces ${\cal F}_I$ and ${\cal F}_{II}$, respectively, and
$E_n$ are assumed to be real and non-negative.}
Note that this statement is almost the same as the following statement 
in statistical mechanics: a canonical state is a maximum of
statistical entropy in the space of states with fixed energy, provided
that the temperature of the canonical state is determined so that the
energy is actually the fixed value.

Note that the expectation value of $E_I$ is equal to the entanglement
energy (\ref{eqn:Eent}), providing that $|n\rangle_I$ and $E_n$ are an
eigenstate and an eigenvalue of the normal-ordered Hamiltonian $:H_I:$
of the subsystem ${\cal F}_I$. 
Hence, for the system of the scalar field, the above general statement
insists that the state (\ref{eqn:HH-state}) is a maximum of the
entanglement entropy in the space of states satisfying SBC, which
corresponds to fixing the entanglement entropy. 
Off course, in this case, the constant $T$ should be determined so
that SBC (\ref{eqn:SBC}) is satisfied.

Returning to the subject, let us prove the general statement. 
(The following proof is the almost same as that given in the Appendix
of Ref.~\cite{Mukohyama1998} for a slightly different statement. 
However, for completeness, we shall give the proof. )

First, we decompose an element $|\psi\rangle$ of ${\cal F}$ as 
%============< EQUATION >==============%
%
\begin{equation}
 |\psi\rangle = \sum_{n,m}C_{nm}|n\rangle_I\otimes |m\rangle_{II}, 
\end{equation}
%======================================%
where the coefficients $C_{nm}$ ($n,m=1,2,\cdots$) are complex numbers 
satisfying $\sum_{n,m}|C_{nm}|^2=1$ and can be considered as matrix
elements of a matrix $C$. Since $C^{\dagger}C$ is a non-negative
Hermitian matrix, it can be diagonalized as
%============< EQUATION >==============%
%
\begin{equation}
 C^{\dagger}C = V^{\dagger}PV,
\end{equation}
%======================================%
where $P$ is a diagonal matrix with diagonal elements $p_{n}$ 
($\ge 0$) and $V$ is a unitary matrix. 
For this decomposition and diagonalization, the entanglement
entropy and the expectation value of the operator $E_I$ are written as 
follows.
%============< EQUATION >==============%
%
\begin{eqnarray}
 S_{ent} & = & -\sum_n p_n\ln p_n, \\
 E_{ent} & = & \sum_{n,m}E_n p_m|V_{nm}|^2, 
\end{eqnarray}
%======================================%
where $V_{nm}$ is matrix elements of $V$. The constraints
$\sum_{n,m}|C_{nm}|^2=1$ and $V^{\dagger}V={\bf 1}$ are equivalent to 
%============< EQUATION >==============%
%
\begin{eqnarray}
 \sum_n p_n & = & 1,\nonumber\\
 \sum_l V^{*}_{ln}V_{lm} & = & \delta_{nm}.
\end{eqnarray}
%======================================%

Next, we shall show that these expressions are equivalent to those
appearing in statistical mechanics in ${\cal F}_I$. 
Let us consider a density operator $\bar{\rho}$ on ${\cal{F}}_I$: 
%============< EQUATION >==============%
%
\begin{equation}
 \bar{\rho} = \sum_{n,m} \tilde{P}_{nm}
                |n\rangle_I\cdot {}_I\langle m|,
\end{equation}
%======================================%
where $(\tilde{P}_{nm})$ is a non-negative Hermitian matrix with unit
trace.  By diagonalizing the matrix $\tilde{P}$ as 
%============< EQUATION >==============%
%
\begin{equation}
 \tilde{P} = \bar{V}^{\dagger}\bar{P}\bar{V},
\end{equation}
%======================================%
we obtain the following expressions for entropy $S$ and an expectation 
value $E$ of the operator
$\bar{E}_I\equiv\sum_n E_n|n\rangle_{I}\cdot{}_{I}\langle n|$. 
%============< EQUATION >==============%
%
\begin{eqnarray}
 S & = & -\sum_n \bar{p}_n\ln \bar{p}_n,\nonumber\\
 E & = & \sum_n \sum_{n,m} E_n\bar{p}_m |V_{nm}|^2,
\end{eqnarray}
%======================================%
where $\bar{p}_n$ is the diagonal elements of $\bar{P}$.
The constraints $\bf{Tr}\bar{\rho}=1$  and 
$\bar{V}^{\dagger}\bar{V}={\bf 1}$ are restated as 
%============< EQUATION >==============%
%
\begin{eqnarray}
 \sum_n \bar{p}_n & = & 1,\nonumber\\
 \sum_l \bar{V}^{*}_{ln}\bar{V}_{lm} & = & \delta_{nm}.
\end{eqnarray}
%======================================%

From these and those expressions, the following correspondence is
easily seen:
%============< EQUATION >==============%
%
\begin{eqnarray}
 S_{ent}	& \leftrightarrow & S,\nonumber\\
 E_{ent}	& \leftrightarrow & E,\nonumber\\
 C^{\dagger}C	& \leftrightarrow & \tilde{P}.
\end{eqnarray}
%======================================%
Hence, a maximum of $S$ in the space of statistical states with a
fixed value of $E$ gives a set of maxima of $S_{ent}$ in the space of
quantum states with a fixed value of $E_{ent}$. (All of them are 
related by unitary transformations in the subspace ${\cal F}_{II}$.) 
Thus, since the thermal state $\tilde{P}_{nm}=e^{-E_n/T}\delta_{nm}$
is a maximum of $S$ in the space of statistical states with a fixed 
value of $E$, $C_{nm}=e^{-E_n/2T}\delta_{nm}$ is a maximum of 
$S_{ent}$ in the space of quantum states with a fixed value of
$E_{ent}$. Here the temperature (or the constant) $T$ should be
determined so that $E$ (or $E_{ent}$) has the fixed value. This
completes the proof of the general statement.

Therefore, for the system of the scalar field, a state of the form
(\ref{eqn:HH-state}) is a maximum of the entanglement entropy in the 
space of quantum states satisfying SBC, provided that the constant $T$ 
is determined so that SBC is satisfied. The value of $T$ is easily
determined as $T=T_H$ by using the well-known fact that the negative
divergence in the Boulware energy density can be canceled by thermal
excitations if and only if temperature with respect to the time $t$ is
equal to the Hawking temperature.

Finally, we obtain the statement that the Hartle-Hawking
state~\cite{Hartle&Hawking} is a maximum of entanglement entropy in
the space of quantum states satisfying SBC since the Hartle-Hawking
state is actually of the form (\ref{eqn:HH-state}) with
$T=T_H$~\cite{Israel1976}.
[Strictly speaking, in order to obtain the Hartle-Hawking state, we
have to take the limit $\alpha\to 0$ (and $L\to\infty$). However, the
following arguments still hold for a finite value of $\alpha$ (and
$L$).] 
The corresponding reduced density matrix is the thermal state with
temperature equal to the Hawking temperature. Therefore, the maximal
entanglement entropy is equal to the thermal entropy with the Hawking
temperature, which is sought in the brick wall model.

%==========================%
%<<<<<   Conclusion   >>>>>%
%==========================%
\section{Conclusion}

In summary the brick wall model seeks the maximal value of
entanglement entropy. 
In other words, the entanglement approach reduces to the brick wall
model when we seek the maximal entanglement entropy .

Our arguments suggests strong connection among three kinds of
thermodynamics: black hole thermodynamics, statistical mechanics, 
and entanglement 
thermodynamics~\cite{Mukohyama1998,MSK1998,D-thesis,MSK1997}. 
It will be interesting to investigate close relations among them in
detail.

%%%%%%%%%%%%%%%%%%%%%%%%%%%%%%%%%%%%%%%%%%%%%%%%%%%%%%%%%%%%%%%%%%%%
%%%%%%%%%%%%%%%%%%%%%%%%%%%%%%%%%%%%%%%%%%%%%%%%%%%%%%%%%%%%%%%%%%%%
% Acknowledgements
%%%%%%%%%%%%%%%%%%%%%%%%%%%%%%%%%%%%%%%%%%%%%%%%%%%%%%%%%%%%%%%%%%%%
%%%%%%%%%%%%%%%%%%%%%%%%%%%%%%%%%%%%%%%%%%%%%%%%%%%%%%%%%%%%%%%%%%%%
\begin{acknowledgments}
The author would like to thank Professors W. Israel and H. Kodama for
their continuing encouragement. 
This work was supported partially by the Grant-in-Aid for Scientific
Research Fund (No. 9809228).
\end{acknowledgments}

%%%%%%%%%%%%%%%%%%%%%%%%%%%%%%%%%%%%%%%%%%%%%%%%%%%%%%%%%%%%%%%%%%%%
%%%%%%%%%%%%%%%%%%%%%%%%%%%%%%%%%%%%%%%%%%%%%%%%%%%%%%%%%%%%%%%%%%%%
% References
%%%%%%%%%%%%%%%%%%%%%%%%%%%%%%%%%%%%%%%%%%%%%%%%%%%%%%%%%%%%%%%%%%%%
%%%%%%%%%%%%%%%%%%%%%%%%%%%%%%%%%%%%%%%%%%%%%%%%%%%%%%%%%%%%%%%%%%%%

\end{document}